\documentclass[12pt,preprint]{aastex}

\begin{document}

\title{Variations in Stellar Clustering with Environment: Dispersed Star Formation
and the Origin of Faint Fuzzies}


\author{Bruce G. Elmegreen}
\affil{IBM Research Division, T.J. Watson
Research Center, P.O. Box 218, Yorktown Heights, NY 10598,
bge@watson.ibm.com}

\begin{abstract}
The observed increase in star formation efficiency with average
cloud density, from several percent in whole giant molecular clouds
to $\sim30$\% or more in cluster-forming cores, can be understood as
the result of hierarchical cloud structure if there is a
characteristic density as which individual stars become well
defined. Also in this case, the efficiency of star formation
increases with the dispersion of the density probability
distribution function (pdf). Models with log-normal pdf's illustrate
these effects. The difference between star formation in bound
clusters and star formation in loose groupings is attributed to a
difference in cloud pressure, with higher pressures forming more
tightly bound clusters. This correlation accounts for the observed
increase in clustering fraction with star formation rate and with
the observation of Scaled OB Associations in low pressure
environments. ``Faint fuzzie'' star clusters, which are bound but
have low densities, can form in regions with high Mach numbers and
low background tidal forces.  The proposal by Burkert, Brodie \&
Larsen (2005) that faint fuzzies form at large radii in galactic
collisional rings, satisfies these constraints.
\end{abstract}


\section{Introduction} \label{sect:intro}

Stars form in concentrations with a range of densities, from star
complexes, OB associations, and T Tauri associations at the low end
to compact clusters and super-star clusters (SSC) at the high end.
The overall structure is usually hierarchical (Scalo 1985;
Feitzinger \& Galinski 1987; Ivanov et al. 1992; Gomez et al. 1993;
Battinelli, Efremov \& Magnier 1996; Bastian, et al. 2005, 2007;
Elmegreen et al. 2006; see reviews in Efremov 1995; Elmegreen et al.
2000, Elmegreen 2005), and this hierarchy continues even inside the
youngest clusters (Testi et al. 2000; Heydari-Malayeri et al. 2001;
Nanda Kumar, Kamath, \& Davis 2004; Smith et al. 2005a; Gutermuth et
al. 2005; Dahm \& Simon 2005; Stanke, et al. 2006; see review in
Allen et al. 2006).  Most likely, the hierarchy in stars comes from
a hierarchy in the gas (St\"utzki et al. 1998; Dickey et al. 2001),
which is the result of turbulence compression and gravitational
contraction that is self-similar over a wide range of scales (see
review in Mac Low \& Klessen 2004). Clusters form in the densest
parts of this gas and lose their initial sub-structure as the
stellar orbits mix (for simulations of star formation in clusters,
see Klessen \& Burkert 2000; Bonnell, Bate, \& Vine 2003; Li, et al.
2004; Tilley \& Pudritz 2004; Klessen et al. 2005;
V\'azquez-Semadeni, Kim, \& Ballesteros-Paredes 2005; Bonnell \&
Bate 2006; Li \& Nakamura 2006).

Hunter (1999) and Ma\'iz-Apell\'aniz (2001) noted that some massive
star-forming regions (which they called ``scaled OB associations'',
or SOBA's) do not form dense clusters while others with the same
total mass do (e.g., the SSC's). We would like to understand this
difference. Obviously the density of the gas is involved, as dense
clusters require dense gas, but the distinction between SSCs and
SOBAs should also be related to the efficiency of star formation,
because clusters forming at low efficiency disperse quickly when the
gas leaves (Lada, Margulis \& Dearborn 1984). At very low
efficiency, stars form individually without passing through an
embedded cluster phase. Recent observations of giant molecular
clouds (GMCs) show star formation at both high and low densities,
with some embedded stars in dense clusters and others more dispersed
(Megeath et al. 2004; J{\o}rgensen et al. 2006, 2007).

Larsen \& Brodie (2000) discovered ``faint fuzzies'' at intermediate
radii in the disks of the S0 galaxies NGC 1023 and NGC 3384, and
suggested they are old clusters with unusually large radii (7-15 pc)
and low densities. They are gravitationally bound because of their
large ages (8-13 Gyr; Brodie \& Larsen 2002), and they are as
massive as SSC's and halo globular clusters. Such clusters appear to
represent an intermediate stage between dispersed and bound star
formation. They appear to be too low in average density to have had
time for core collapse and envelope expansion as in models of
globular clusters by Baumgardt et al. (2002).  Burkert, Brodie \&
Larsen (2005) suggested they formed in a collisional ring
interaction between two galaxies.

What determines the relative proportion of dispersed and clustered
star formation? Larsen \& Richtler (2000) showed that clustering on
a galactic scale, measured as the fraction of uv light in the form
of massive young clusters, increases as the star formation rate
increases. This could be the result of a selection effect if
starbursts are active for less than a cluster destruction time. On
the other hand, the clustering fraction could depend on pressure.
Higher pressure makes the cool phase of gas denser, which promotes
more clustering, and locally high pressures trigger star formation
on the periphery of GMCs, making clusters in the dense gas (e.g.,
Comer\'on, Schneider, \& Russeil  2005; Zavagno et al. 2006). The
Larsen \& Richtler correlation could then follow from the mutual
correlation between pressure and star formation rate with gas column
density. Faint fuzzies are a counter example, however: they formed
bound but presumably at low pressure to have such low central column
densities now. Can all of these clustering types be understood as a
continuum of properties in a universal physical model?

There have been several attempts to explain the difference between
clustered and distributed star formation based on numerical
simulations. Klessen, Heitsch, \& Mac Low (2000) suggested that
clusters form in non-magnetic gas when the turbulence driving scale
is large. Heitsch, Mac Low \& Klessen (2001) noted that non-magnetic
turbulence driven on large scales produces a clustered collapse,
while magnetic turbulence in supercritical clouds produces a more
distributed collapse. Mac Low (2002) suggested that stars form in
clusters when there is no turbulent support and they form disbursed
when there is. V\'azquez-Semadeni, Ballesteros-Paredes, \& Klessen
(2003) suggest this transition from no global turbulent support to
support corresponds to an increase in the Mach number and a decrease
in the sonic scale, which is the length where the size-linewidth
relation gives a Mach number of unity. Large sonic scale (low Mach
number) corresponds to a sonic mass larger than the thermal Jeans
mass, which means a lack of global support and the formation of a
cluster. Low sonic scale corresponds to the dispersed formation of
stars, one for each tiny compressed region where the mass exceeds
the local thermal Jeans mass. Li, Klessen \& Mac Low (2003)
suggested that the equation of state determines the stellar
clustering properties: soft equations produce dense clusters while
hard equations produce isolated stars. These suggestions all apply
to initially uniform media. External compression of a cloud into a
massive dense core can also make a cluster; most young clusters are
in high-pressure regions like OB associations.

Simulations of turbulent media produce stars in compressed regions
that act as seeds for the small-scale gravitational collapses that
follow (V\'azquez-Semadeni, Ballesteros-Paredes, \& Klessen 2003;
Clark \& Bonnell 2005). These simulations also have probability
distribution functions (pdfs) for density that are either log-normal
or log-normal with a power-law tail at high density, especially when
self-gravity is important (e.g., Li, Klessen \& Mac Low 2003).  The
efficiency of star formation is then proportional to the fraction of
the gas in a dense form. Here we examine variations in this fraction
as functions of average density and velocity dispersion, and as a
function of the local density inside a cloud. The efficiency is
taken to be the ratio of the stellar mass to the gas mass during a
complete star-forming event. It generally increases with cloud
density from a few percent in GMCs (Williams \& McKee 1997) to
several tens of percent in cluster-forming cores (e.g., Lada \& Lada
2003). It may reach $\sim50$\% or more inside the densest
star-forming cores. We explain this increase as a result of
hierarchical structure, regardless of the dynamics and mechanisms of
star formation, and we show that for log-normal or similar density
pdf's, as expected in turbulent media, the mass fraction of regions
with high efficiency increases with the Mach number and,
independently, with the average density. This result may explain the
Larsen \& Richtler (2000) correlation as well as the observed
variations in clustering properties with pressure. We also show that
at high Mach number, bound clusters can form with relatively low
average densities, thereby explaining faint fuzzies. These are all
consequences of star formation in the dense cores of clouds that are
structured by turbulence. They result primarily from the geometry of
the gas, which is somewhat universal, and should be nearly
independent of the gas dynamics or the strength of the magnetic
field.

We make an important assumption that gravitational contraction and
star formation can occur in regions that are either larger or
smaller than the sonic scale. This means we assume that contraction
to one or more stars can occur in a supersonically turbulent region.
V\'azquez-Semadeni, Ballesteros-Paredes, \& Klessen (2003) suggest
that if a cloud is supported by turbulence, then only regions
smaller than a sonic scale and more massive than the thermal Jeans
mass are unstable to form stars. Padoan (1995) was the first to
consider this condition. However, clouds are probably not supported
for any significant time by turbulence, and even if they were, it is
only necessary that a clump mass exceed the turbulent Jeans mass for
self-gravitational forces to exceed inertial forces. GMCs for
example, have comparable self-gravitational and turbulent energy
densities and yet are much larger than the sonic length. Our
assumption is contrary to that in Krumholz \& McKee (2005), who
assume the same as V\'azquez-Semadeni et al..  We are consistent
with McKee \& Tan (2003), however, as they consider the collapse of
a highly turbulent core to make a massive star. Saito, et al.
(2006), for example, observe star formation in massive turbulent
cores. Thus the sonic length should not provide a threshold for star
formation.

Our primary condition for star formation is that the gas density
exceed some fixed value, taken here to be $10^5$ cm$^{-3}$. This is
the density at which HCN gives a nearly constant star formation rate
per unit gas mass (Gao \& Solomon 2004a,b; Wu et al. 2005) and at
which a variety of microscopic processes conspire to shorten the
magnetic diffusion time (Elmegreen 2007). Regions with this density
should have a wide range of Mach numbers but a nearly universal
efficiency, according to the Gao \& Solomon and Wu et al.
observations. The density of $\sim10^5$ cm$^{-3}$, when converted to
5900 M$_\odot$ pc$^{-3}$, is also typical for star clusters, as most
of those surveyed by Tan (2007), which span a factor of $10^5$ in
mass, have about this average density. The cluster density equals
the gas density times the efficiency, and the efficiency has to
exceed $\sim10-30$\% for a bound cluster to form. Thus, the gas
density for both cluster formation and high efficiency appears to be
around $10^5$ cm$^{-3}$ or, possibly, $10^6$ cm$^{-3}$. We note that
the long timescale derived for HCN gas as the ratio of the total HCN
mass divided by the total star formation rate (Gao \& Solomon 2004b;
Wu et al. 2005) is not the duration of star formation in any one
place, but is the HCN consumption time. As long as a new HCN region
is formed somewhere each time an old HCN region disperses, the HCN
consumption time can be long even when each region of star formation
lasts for a short time (see Elmegreen 2007 for a discussion of star
formation timescales).  As a result, the average efficiency of star
formation in any one HCN region can be moderately large even if the
average HCN consumption time is long. An average efficiency of
$\sim5-10$\% would be reasonable considering that most star-forming
regions leave unbound clusters after the gas leaves (Lada \& Lada
2003), most regions are observed at half their total ages, and star
formation typically accelerates over time (Palla \& Stahler 2000).
With constant acceleration, only $\sim1/4$ of the total stars form
in the first 1/2 of the total time. The efficiency that determines
whether a bound cluster will remain is taken here to be 14.4\% (see
below), and the peak efficiency in a single-star core is taken to be
50\% at $10^5$ cm$^{-3}$ density. These values are uncertain and are
used here only to illustrate how cluster formation might scale with
the velocity dispersion and density of the cool component of the
interstellar medium.

\section{Cluster Formation in a Hierarchical ISM: Bound and Unbound
Clusters} \label{sect:clus}

The hierarchical structure of interstellar gas implies that the mass
fraction of star-forming cores at high density $n_c$ increases in
regions of the cloud that have a higher average density. This
density dependence may be illustrated with a simple model. Consider
a local star formation rate proportional to $\epsilon(\rho)
\rho\left(G\rho\right)^{1/2}$ for $\rho<\rho_{c}$ for efficiency
$\epsilon$. In this paper, mass density will be denoted by $\rho$
and molecule density by $n$. If we denote the galactic average
quantities by a subscript ``0'', then the Kennicutt-Schmidt star
formation rate is analogous to $\epsilon_0 \rho_0 (G\rho_0)^{1/2}$
for $\epsilon_0\sim0.012$ (Elmegreen 2002a). The same galactic rate
would be obtained for observations at any other density, provided
the efficiency and volume filling factor at that density are
properly scaled. Thus we write:
\begin{equation}\epsilon_0 \rho_0 (G\rho_0)^{1/2} =
\epsilon(\rho) \rho (G\rho)^{1/2}f_V(\rho) = \epsilon_{c}
\rho_{c}(G\rho_{c})^{1/2}f_V(\rho_{c}),\end{equation} where
$f_V(\rho)$ is the fraction of the volume having a density larger
than $\rho$. This expression with the volume fraction can be
converted to one with the mass fraction using the relations $\rho
f_V(\rho) = \rho_0f_M(\rho)$ and $\rho_{c} f_V(\rho_{c}) =
\rho_0f_M(\rho_{c})$, where $f_M(\rho)$ is the fraction of the mass
having a density larger than $\rho$.  Subscripts {\it c} denote gas
at the threshold density for a constant efficiency. As a result,
\begin{equation}\epsilon(\rho) = \epsilon_{c}
\left(\rho_{c}/\rho\right)^{1/2} \left[f_M\left(\rho_{c}\right)/
f_M\left(\rho\right)\right].\label{eq:eps}\end{equation}

\begin{figure}\epsscale{.75}
\plotone{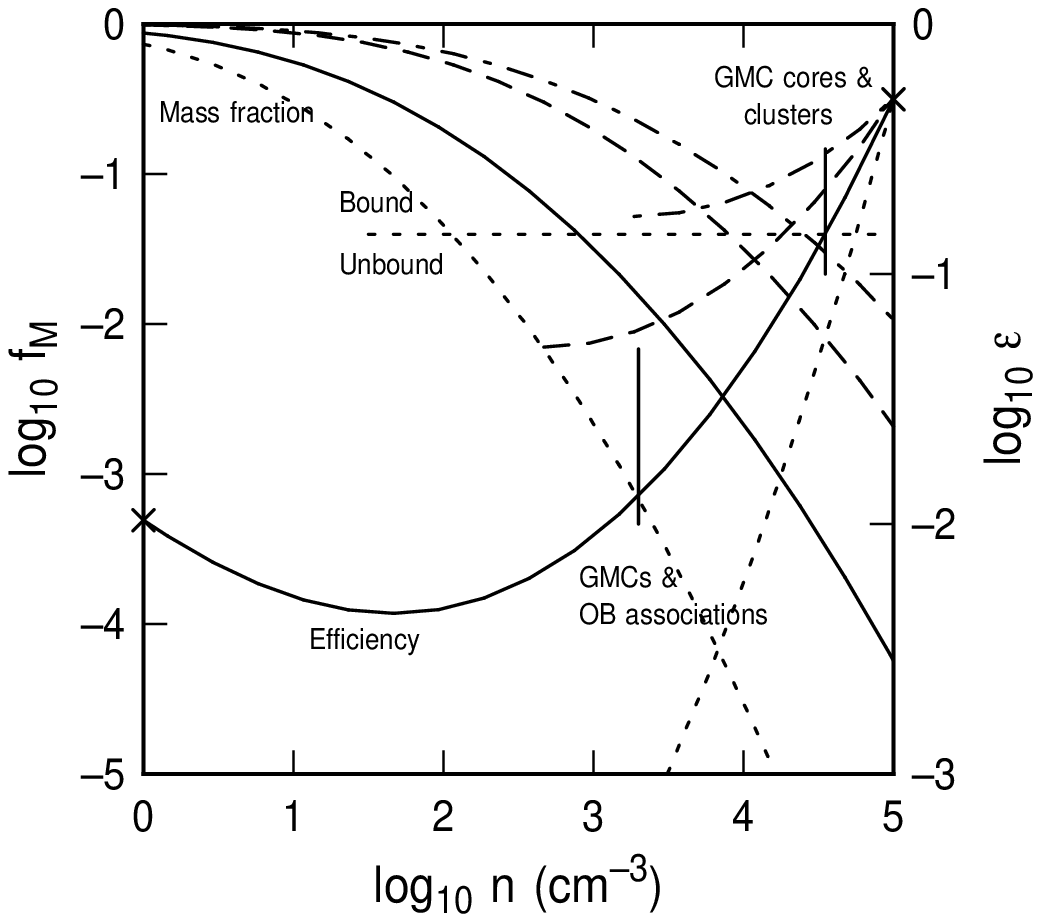}\caption{Left-hand axis. and downward sloping lines:
The fraction of the mass in a turbulent region, $f_M$, modeled with
a log-normal density pdf, that has a density greater than the value
on the abscissa. Right-hand axis and upward sloping lines: the local
efficiency $\epsilon$ of star formation as a function of average
local density. Observed efficiencies for OB associations and for
clusters are shown by the vertical lines. The efficiency is taken to
be the fraction of the cloud mass having a density greater than
$10^5$ cm$^{-3}$, multiplied by the efficiency of star formation at
the threshold, which is 50\% here.  The different line types
correspond to normal galaxies (solid line, $[\sigma,n_0]=[2.3,1]$),
starbursts (dashed $[2.3,10]$ and dot-dashed $[2.56,10]$) and
quiescent regions (dotted $[1.98,0.5]$). Here $\sigma$ is the
dispersion of the log-normal pdf and $n_0$ is the average density.
The mass fraction of gas with large $\epsilon$ ($>0.144$ in this
example) is larger in starburst regions than it is in normal
galaxies; it is lower in quiescent regions. This implies that
starbursts should form most of their stars in bound clusters, while
quiescent regions should form most of their stars in unbound
associations. The figure also shows that the minimum density for
bound clusters decreases with increasing Mach number; faint fuzzies
may form as an extreme example of this.}\label{fig:fractio3}
\end{figure}

Mass fractions and efficiencies for log-normal density pdfs are
shown in Figure \ref{fig:fractio3} for $n_c=10^5$ cm$^{-3}$. The
decreasing lines are the mass fractions $f_M(\rho)$ at densities
larger than the value on the abscissa, and the increasing lines,
which are $\epsilon(\rho)$, are $\epsilon_c$($=0.5$) times the mass
fractions of the clumps of density $\rho_c$ inside regions of
average density $\rho$. For this log-normal case, the normalized pdf
is
\begin{equation}
P(x)={{1}\over{\left(2\pi\right)^{1/2}\sigma}}
e^{-0.5\left(x^2-x_p^2\right)/\sigma^2}\end{equation} where
$x=\ln\rho$ and $x_p=\ln\rho_p$ at the pdf peak
($\rho_p=\rho_0e^{-\sigma^2/2}$ for average $\rho_0$). The pdf for
equal increments in density is $P^\prime\left(\rho\right)=P(x)/\rho$
and the mass fraction greater than some density is
$f_M(\rho)=\int_\rho^\infty \rho P^\prime\left(\rho\right)d\rho$.

The density pdfs in the figure have dispersions and average
densities of 2.3 and 1 cm$^{-3}$ for the solid line, 2.3 and 10
cm$^{-3}$ for the dashed line, 2.58 and 10 cm$^{-3}$ for the
dot-dash line, and 1.98 and 0.5 cm$^{-3}$ for the dotted line. The
first of these dispersions comes from galaxy simulations in Wada \&
Norman (2001) and from the low density case in Wada \& Norman
(2007); the density of 1 cm$^{-3}$ is typical for normal galaxy
disks. The second has a higher density, similar to the inner parts
of galaxies, and the third has the same high density and a Mach
number that is twice the effective value in Wada \& Norman (2001),
as determined from the relation between dispersion and Mach number
in Padoan \& Nordlund (2002), which is
$\sigma=\left(\ln\left[1+0.25M^2\right]\right)^{1/2}$ for Mach
number $M$.  These latter two cases represent moderately strong
starburst regions. The fourth case has a Mach number that is half
the Wada \& Norman value and a density that is also small. This case
applies to quiescent regions with low-intensity star formation, such
as the outer parts of galaxies, parts of dwarf galaxies, and low
surface brightness galaxies.  Wada \& Norman (2007) simulate
self-gravitating disks and find log-normal density pdfs with
$\sigma$'s that increase from 2.4 to 3.0 as the initial density
(within 10 pc of the midplane) increases from 5 to 50 M$_\odot$
pc$^{-3}$. Our $\sigma$ are in this range.  The equilibrium peak
densities in the Wada \& Norman simulations are all about $\sim2$
cm$^{-3}$, which is also comparable to the values used here.

The log-normal nature of the assumed density pdf is not critical to
the conclusions of this paper. Any density pdf with an extended tail
at high density and a breadth that increases with Mach number will
give the same correlations between bound cluster fraction, Mach
number, and pressure. Similarly, any pdf that has a smaller slope
near the average density than at high density (like a log-normal)
will give our additional dependence of the bound cluster fraction on
the average density. The main point is that when the slope of the
density pdf is relatively shallow near the critical density for star
formation, bound clusters can form with a wide range of average
densities. These types of pdfs will also produce faint fuzzies in
the limit of high dispersion as these clusters result from highly
efficient star formation at a low average cloud core density. The
absolute calibration of the density pdf is also not critical to our
model. We could equally well consider pdfs for cluster-forming cores
where the average density is $\sim10^2$ cm$^{-3}$ and the critical
density for ``final'' collapse is $\sim10^7$ cm$^{-3}$. The
essential points are that (1) for all of these pdfs, the mass
fraction of the ``final'' collapsed cores, and therefore the
efficiency of star formation in a particular region,
$\epsilon(\rho)$, increases with the average density of the region
(an implication of hierarchical structure), and (2) the density pdf
is more or less flat at the critical density of star formation,
depending on some physical variable, taken here to be the Mach
number and/or average density.  Wada \& Norman (2007) point out that
there is no single Mach number in their simulations but a range of
values, and still the density pdf is log-normal. They find the
primary dependence of $\sigma$ to be on the initial disk density.
The origin of this $\sigma$ is not important here, only the effect
that variable $\sigma$ and $\rho_0$ have on the mass fraction of
dense gas.

According to the equations which define $\epsilon(\rho)$ for the
log-normal model, there is a minimum value in each $\epsilon(\rho)$
curve that occurs where $d\ln f_M/d\ln \rho=-0.5$. At lower density,
the curve $f_M(\rho)$ flattens and $\epsilon(\rho)$ is dominated by
$\rho^{-1/2}$ so it increases with decreasing $\rho$. At higher
density, $\epsilon(\rho)$ is dominated by the Gaussian tail in
$f_M(\rho)$ so it increases with increasing $\rho$. The low density
behavior of $\epsilon(\rho)$ is not realistic because the low
density part of $f_M$ is not likely to remain a log-normal. This
part corresponds to the low density intercloud medium and the
physical processes there are different than in dense gas. It may be
controlled by supernova cavities, for example. In Wada \& Norman
(2001), the low density part of $f_M$ was not a log-normal. Thus we
consider only the increase in $\epsilon(\rho)$ with $\rho$ to be
characteristic of dense gas involved with star formation.

For each of the cases considered in Figure 1, the efficiency
increases with density because hierarchical structure gives $n_{c}$
cores a higher filling factor at higher average densities. The mass
fraction decreases with density because only a small fraction of the
matter is dense. For the fiducial case (solid line), an average
galactic efficiency, $\epsilon(\rho_0)=0.01$ (e.g., Kennicutt 1998),
is indicated by the cross on the left and a typical efficiency for a
dense star-forming core, $\sim50$\% at $\rho_{c}\sim10^5$ $m(H_2)$
cm$^{-3}$, is indicated by the cross on the right. The vertical
lines show typical ranges for efficiencies and densities in OB
associations ($\epsilon\sim1$\% to 5\% at $n\sim10^{3.3}$ cm$^{-3}$)
and in the cluster-forming cores of OB associations
($\epsilon\sim10$\% to 30\% at $n\sim10^{4.5}$ cm$^{-3}$).

The approximate value of the efficiency separating clusters that end
up mostly bound after gas dispersal from clusters that become mostly
unbound ($\epsilon_{cluster}$) is placed in the figure at
$\epsilon(\rho)\sim0.144$, which is the value of $\epsilon(\rho)$ in
the fiducial case at a gas density of $n=10^{4.55}$ cm$^{-3}$. This
efficiency is shown by the dotted horizontal line; its exact value
is not important here. The efficiency for bound cluster formation
depends on the rate of gas dispersal and the initial velocities of
the stars. Slower gas dispersal rates and smaller initial speeds
compared to virial require lower efficiencies for boundedness (e.g.,
Verschueren 1990; Lada \& Lada 2003; Boily \& Kroupa 2003; Goodwin
\& Bastian 2006).

The basic trends in Figure \ref{fig:fractio3} illustrate the
important differences between cluster formation in various
environments. First, the threshold efficiency occurs at lower
average densities for starburst pdfs and at higher average densities
for quiescent pdfs. That is, the density $\rho$ at which
$\epsilon(\rho)=0.144$ in the figure is smaller than $10^{4.55}$
$m(H_2)$ cm$^{-3}$ in the first case and larger than $10^{4.55}$
$m(H_2)$ cm$^{-3}$ in the second. When the average density for
$\epsilon>\epsilon_{cluster}$ is much lower than the inner core
density $n_{c}$, a high fraction of the gas mass can form stars with
$\epsilon>\epsilon_{cluster}$. Long-lived clusters should be easier
to form in this case. Because this is the starburst limit, the
result is in qualitative agreement with the observations by Larsen
\& Richtler (2000).

\begin{figure}\epsscale{.9}
\plotone{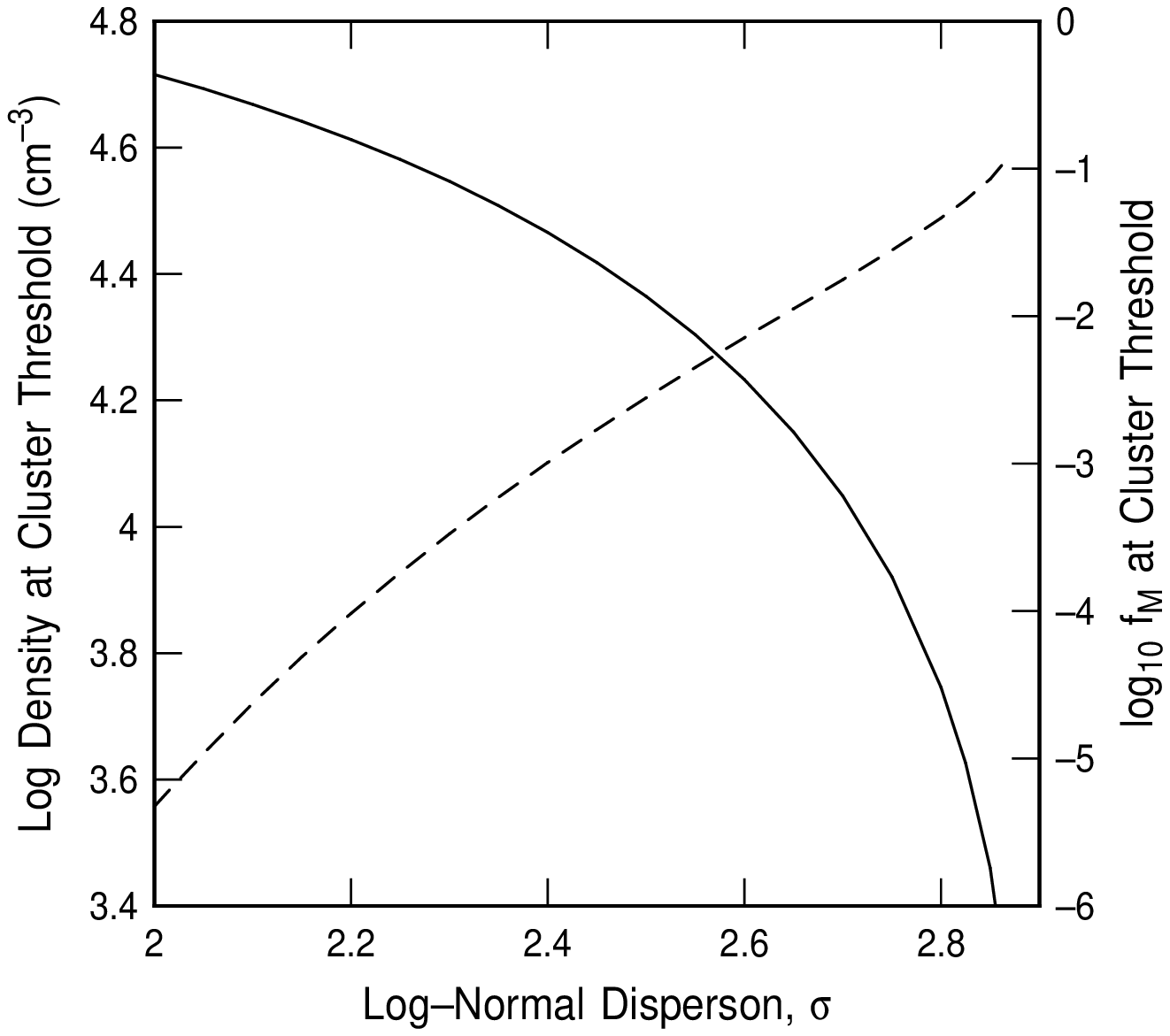}\caption{(Left-hand axis, solid line) The density at
a fixed threshold efficiency for cluster formation,
$\epsilon(\rho)=\epsilon_{cluster},$ is shown as a function of the
dispersion $\sigma$ of the density pdf for cloud structure.
(Right-hand axis, dashed line) The mass fraction of gas at a density
greater than the threshold density for cluster formation, versus
$\sigma$. The average gas density is taken to be 1 cm$^{-3}$,
consistent with the solid lines in Fig. 1. The trends observed here
arise for a wide range of threshold efficiencies, so the precise
value is not important; $\epsilon_{cluster}=0.144$ from Fig. 1 was
used.  The figure indicates that ISM regions with higher dispersions
for their density pdfs (e.g., higher Mach numbers) have lower
relative density thresholds for bound cluster formation and higher
fractions of the total mass forming bound
clusters.}\label{fig:epsilon}\end{figure}

Figure \ref{fig:epsilon} shows this result by plotting as a solid
line the density at which $\epsilon(\rho)=0.144$ versus the
dispersion $\sigma$ of the log normal density pdf, for an average
density of 1 cm$^{-3}$. The threshold density value of $10^{4.55}$
cm$^{-3}$ when $\sigma=2.3$ from Figure 1 is one point on the curve.
Higher dispersions correspond to lower threshold densities for bound
cluster formation. Figure \ref{fig:epsilon} also shows as a dashed
line the fraction of the mass, $f_M$, at densities greater than the
threshold density as a function of the dispersion.  Larger
dispersions lead to a larger fraction of the mass in bound clusters,
mostly because the threshold density decreases and more gas mass is
above the threshold.

When the average density for $\epsilon>\epsilon_{cluster}$ is
comparable to $n_{c}$, the mass fraction of gas with highly
efficient star formation is very low. This case applies to low Mach
numbers or low pressures and may explain why in some regions like
the giant OB association NGC 604 in M33 there is a lot of star
formation and thousands of young stars, but few bound clusters
(Ma\'iz-Apell\'aniz 2001).

We propose that the difference between the formation of clusters and
the formation of loose OB associations or stellar groupings of the
same total mass and age depends mostly on the ratio of the
characteristic density for stellar core formation, $\rho_c$, to the
density at which $\epsilon(\rho)=\epsilon_{cluster}$. High pressure
gas has a high density ratio and should form a high proportion of
stars in bound clusters. Low pressure gas has a density ratio near
unity and should form a low proportion of stars in bound clusters
with most forming in loose OB associations.

Some GMCs in the solar neighborhood have substantial populations of
protostars outside the dense clusters (Megeath et al. 2004).
J{\o}rgensen et al. (2006, 2007) found that 30-50\% of the
protostars in Perseus are outside the two main clusters.  This is to
be expected in hierarchical clouds because not all of the
star-forming clumps are in the cluster-forming cores. In a high
pressure region, a hypothetical Perseus-like collection of
protostars would form at a higher average density, putting the
individual stars closer together and giving them a higher fraction
of the total cloud mass where they form. As a result, they would be
more tightly bound to each other when the gas leaves. In a low
pressure region, the cores would be further apart with relatively
more intercore gas, and they would be more likely to disperse into
the field along with the gas. In both cases, a fixed fraction of the
turbulence-compressed cores could form stars, but the mass fraction
represented by these cores can be high or low, depending on the Mach
number and density. The conditions in the Perseus cloud are
apparently intermediate between these two limits.

Johnstone et al. (2004) found that 2.5\% of the $\rho$ Ophiuchus
cloud mass is in dense sub-mm cores, while Kirk et al. (2006) found
that 0.4\% of the mass in the Perseus cloud is at $A_V>10$ mag.
Enoch et al. (2006) observed $<5$\% of the Perseus cloud mass in 1.1
mm cores.  These small fractions are consistent with the turbulent
fragmentation model of Figure \ref{fig:fractio3}, as shown by the
decreasing $f_M(\rho)$ lines; recall that $f_M$ is the fraction of
the total mass at density larger than $\rho$ (while $\epsilon(\rho)$
is the fraction of the mass having a density greater than fixed
$\rho_c$ in regions of density $\rho$). At high density, $f_M(\rho)$
is low while $\epsilon(\rho)$ is high. If each sub-mm core evolves
on its own dynamical time, and the surrounding cloud does the same
as it forms new cores, then the cloud will have a number of
core-forming events equal to the square root of the ratio of
densities, which is $\sim10$. Thus the net efficiency of core
formation can build up to several tens of percent over time, and the
final efficiency of star formation might be $\sim10$\% or more,
considering the efficiency inside each sub-mm core and the
likelihood that some cores will disperse without forming a star
(e.g. V\'azquez-Semadeni et al. 2005).  A large number of core
crossing times is not necessary for this to happen. The increasing
$\epsilon(\rho)$ curves in Figure \ref{fig:fractio3} illustrate
qualitatively a second point made by these authors, and by
J{\o}rgensen et al (2007) as well for Perseus, that the mass
fraction of cores becomes high in the generally denser regions of
the cloud; i.e., at high extinction. They find an $A_V$ extinction
threshold for the occurrence of sub-mm cores equal to 5 to 7 mag (15
mag in Johnstone, et al). Figure \ref{fig:fractio3} suggests that it
is not the extinction, per se, which produces this correlation, but
the hierarchical structure of clouds.

A second result from Figures \ref{fig:fractio3} and
\ref{fig:epsilon} is that a bound cluster can have a lower average
density when the Mach number or ambient density is higher, i.e.,
when the pdf is flatter or shifted to higher $\rho$. This suggests
an explanation for faint fuzzies: they are normal clusters that form
at the limiting lower density for boundedness in high pdf-dispersion
regions. The lowest density for a bound cluster is lower when the
$\epsilon(\rho)$ curve in Figure \ref{fig:fractio3} is flatter at
$\rho_c$. This makes the density at $\epsilon(\rho)=
\epsilon_{cluster}$ lower and corresponds to a broader pdf
dispersion $\sigma$ in Figure \ref{fig:epsilon}. For an isothermal
gas, a broader dispersion corresponds to a higher Mach number. Faint
fuzzies also require low background tidal forces or they would have
broken apart by now. Usually, high Mach numbers correspond to high
critical tidal densities because both occur in starburst regions.
But faint fuzzies require an odd combination: high Mach numbers (or
broad pdf-dispersions) and low tidal densities.  This is possible in
highly shocked clouds that occur in remote regions of a galaxy. For
example, during a galaxy collision, shocked clouds in the outer
regions (or direct ISM impacts in the outer regions) can have Mach
numbers approaching the pair orbital speed, but at these locations,
the local tidal forces are fairly low. This result fits well with
the model of faint fuzzies by Burkert, Brodie \& Larsen (2005), who
suggested they formed in galaxy collisional rings. Such rings are
indeed highly shocked regions in the outer parts of galaxies, where
the tidal density is fairly low. Burkert et al. did not explain how
such conditions would produce faint fuzzies, however. Here, they
result naturally in a turbulent ISM model as a consequence of the
flattening of the density pdf at higher Mach number.

\section{Multiple Star Formation Events inside the Hierarchies}

The observed hierarchies of star formation, such as subgroups inside
OB associations, which, in turn, are inside star complexes, have the
property that the duration of star formation increases approximately
as the square root of size (Efremov \& Elmegreen 1998) and is always
in the range of 1 to 2 crossing times (Elmegreen 2000, 2007).  The
smaller regions come and go relatively quickly while the larger
regions form stars, and there are many small regions within the
spatial bounds and during the total time span of the large region.
When the duration of star formation is always about a crossing time,
then the number of events at the density $\rho$ is proportional to
$\left(G\rho\right)^{1/2}$. These small-scale events are not all at
the same place, but they move around inside the large scale event as
clouds get dispersed and cloud envelopes get triggered.

For each large-scale event with cloud mass $M$ ($\sim10^7$ M$_\odot$
in main galaxy disks) and average density $\rho_0$, the total mass
of all the gas in star forming events at any one time at a density
greater than $\rho$ is $Mf_M\left(\rho\right)$ according to the
definition of $f_M$ in the previous section. The star formation
efficiency in each of these high density regions is
$\epsilon\left(\rho\right)$. The number of events of star formation
at the density $\rho$ during the lifetime of the larger-scale cloud
is $\left(G\rho\right)^{1/2}/\left(G\rho_0\right)^{1/2}$, according
to the previous paragraph.  The product of these three terms is the
total stellar mass formed in all the fragments of the large-scale
cloud that ever had a density $\rho$. This product is
$Mf_M(\rho)\epsilon(\rho)\left(\rho/\rho_0\right)^{1/2}$.
Substituting for $\epsilon(\rho)$ from equation \ref{eq:eps} and
using the identity
$f_M\left(\rho_c\right)=f_V\left(\rho_c\right)\rho_c/\rho_0$ from
the discussion just before equation \ref{eq:eps}, all of the terms
involving density cancel and we get simply $\epsilon_0M$. This
result confirms the basic nature of hierarchical star formation: the
mass of stars forming in each level of the hierarchy is the same and
equal to the total mass of stars formed. This is a property of
hierarchies because stars which form in the densest level are the
same stars as those which form in the lower-density levels that
contain the densest levels.  The important point for star formation
is that this hierarchy is in both space and time. The answer comes
out correctly only when the multiple events of star formation at
each density are considered.

\section{Stellar Mixing and the Formation of Clusters}

If all star-forming regions last for about a crossing time before
they disperse and reform into other star-forming regions inside the
same larger-scale regions (Elmegreen 2000, 2007), then there is an
equal opportunity for stars on all scales to mix in their orbits and
become smooth clusters. This means that the oldest stars in all
groupings, ranging from embedded clusters to giant star complexes,
should be fairly well mixed while the youngest stars in these
groupings should still be hierarchical. The scale dependence of the
mixing fraction has never been observed systematically, but general
observations suggest it is approximately true. For example, the
youngest objects in the Orion nebula, the proplyds, still cluster
near $\theta^1$ Ori C while the slightly older stars are more
dispersed (Smith et al. 2005b; Bally et al. 2005). Similarly, the
youngest objects in Gould's Belt, which could be the nearest star
complex, are clustered into the local OB associations such as Orion,
Perseus, and Sco-Cen, while the oldest objects, the Pleiades and
Cas-Tau associations, are now dispersed into stellar streams (e.g.,
Elias, Alfaro \& Ca\~no 2006). The large regions never get
self-bound, unlike the small regions, because the efficiency of star
formation is always small on large scales, as shown in Section
\ref{sect:clus}. Still these large regions mix through the action of
random stellar drift.

Galactic shear pulls apart the large-scale groupings over tens of
millions of years, which is the crossing time. The shear time is
independent of size and equal to the inverse of the Oort A constant,
while the crossing time increases with size approximately as the
square root. This means there is a sufficiently large size where the
crossing time equals the shear time and at larger sizes, the
crossing time exceeds the shear time. This critical size is always
about the scale height in the galaxy for Toomre $Q\sim1$ (Elmegreen
\& Efremov 1996). Larger regions look like star steams or flocculent
spiral arms because shear dominates, while smaller regions look
clumpy like star complexes and OB associations because the internal
dynamics dominates.

\section{Discussion}

Stars form bound clusters where the total efficiency is high.  The
efficiency should be proportional to the mass fraction of a cloud in
the form of dense cores where the individual stars form. This mass
fraction increases with the cloud density in hierarchical clouds. It
also increases with the Mach number of the turbulence because
stronger shocks at higher Mach numbers compress the gas to a wider
range of densities. As a result, a log-normal density pdf becomes
flatter below the threshold density of star formation, and this
means the mass fraction is higher at each density below this value.
When the Mach number is high (really, when the dispersion of the
density pdf is high), the efficiency is high even at a fairly low
average cloud density, and so a high fraction of star formation ends
up bound. An extreme example of this trend is the type of cluster
called a faint fuzzie. We propose that faint fuzzies form in
moderately low density clouds with moderately high Mach numbers. The
galactic tidal ring environment proposed by Burkert et al. (2005) is
an example of a region that would have such clouds.

The mass fraction of clumps at a particular high density of star
formation also increases with the average density of the ISM because
then the whole density pdf shifts toward higher values. The
combination of high densities and high Mach numbers, characteristic
of starburst regions, makes for a high fraction of star formation in
bound clusters.

On the other hand, relatively low Mach numbers and/or low densities
should produce stars in a more dispersed way, as in the low-density
regions of molecular clouds or in low surface brightness galaxies
and regions of galaxies. This combination of parameters corresponds
to a low ISM pressure, so we infer that low pressure regions, which
means those with a low gas column densities and low star formation
rates per unit area, should produce relatively fewer bound clusters
and relatively more unbound associations.

There is a physical explanation for the trends discussed here.
Consider a moderately low density cloud with a low turbulent Mach
number. The compressions inside that cloud will be modest and most
of them will not reach a level of density or enhanced magnetic
diffusion rate that allows gravitational collapse before the
compression ends.  Then few stars will form and the efficiency will
be low (unless the cloud continues to makes these weak compressions
for a very large number of crossing times, which seems unlikely).
Now consider this same cloud with the same size and average density
but with a higher Mach number (this will require a higher ISM
pressure). The stronger compressions will more easily produce
high-density cores in which gravity is important and magnetic
diffusion is fast. More stars will form and the efficiency will be
high. The only difference between these two examples is the Mach
number, which affects the range of densities in the compressed
regions. In terms of the density pdf, higher Mach numbers produce a
broader pdf at the same average density.  In terms of star formation
in bound clusters, high pressure regions having clouds with high
Mach number produce a higher fraction of their stars in bound
clusters.

The model predicts that young bound clusters in starburst regions,
or in regions of high ISM pressures or Mach numbers, will have a
wider range of densities than bound clusters in more quiescent,
low-pressure regions. This is partly because the cloud densities
should be higher in starburst regions, so the resulting clusters can
be denser overall, but it is also because the lower-density parts of
starburst clouds can form stars with a high efficiency, leaving
bound clusters when the gas leaves rather than dispersed OB
associations. As there is generally more mass at low density than at
high density, the net distribution of cluster density could shift
toward lower values when the ISM density pdf is broad. According to
the model, this density shift is relative to the mean ISM density,
and it should be measured only before significant cluster expansion.
Low surface brightness galaxies should make a preponderance of
unbound OB associations rather than bound clusters, and the young
clusters they form should have a relatively narrow range of
densities compared to normal.

\section{Summary}

Hierarchical structure in both gas and young stars is evident from
fractal analysis, correlation functions, and power spectra (see
references in Sect. I) and is probably the result of turbulence
compression and self-gravity. Stars form in the densest part of the
gas. In regions where the individual star-forming clumps represent a
high fraction of the mass, the cluster that forms has a good chance
of remaining self-bound after the gas leaves. Such regions typically
have densities of $\sim10^{4}$ cm$^{-3}$ or more in the solar
neighborhood. We suggest here that star formation at a high average
density like this automatically has a high efficiency because the
individual star-forming clumps are only an order of magnitude or two
denser. This follows entirely from the hierarchical ISM. In such a
medium, clumps of any density cluster together with clumps of a
similar density, and the mass fraction of these clumps increases as
the average density of the surrounding region increases.  As a
result, low density clouds like GMCs form stars with a low total
efficiency and high density cloud parts like GMC cores form stars
with a high total efficiency, leaving bound or marginally bound
clusters.

The density where the efficiency first becomes high decreases as the
Mach number of the turbulence increases and as the average density
of the ISM increases. Consequently, high pressure regions should
place a higher fraction of their stars in bound clusters, while low
pressure regions should preferentially make unbound stellar
groupings. Regions with moderately low density and moderately high
Mach number should be able to make faint fuzzies, which are
low-density bound clusters. Regions with high densities and low Mach
numbers should make extremely dense clusters.

The masses of the high density regions and of the star clusters that
form in them have not been specified in this derivation of
$\epsilon(\rho)$. These masses should vary over a wide range and be
distributed approximately as $dN/dM\sim M^{-2}$ because of the
hierarchical nature of the gas (Fleck 1996; Elmegreen \& Efremov
1997; Elmegreen 2002b). Thus bound and dispersed stellar groupings,
as well as faint fuzzies, should have the usual $M^{-2}$ mass
distributions.

\acknowledgements
My thanks to Mark Gieles for a helpful referee's
report.

\clearpage

\begin{thebibliography}{}

\bibitem[]{637} Allen, L., Megeath, S. T., Gutermuth, R., Myers,
P. C., Wolk, S., Adams, F. C., Muzerolle, J., Young, E., \& Pipher,
J. L. 2007, in Protostars and Planets V, ed. B. Reipurth, D. Jewitt,
\& K. Keil, (Tucson, Univ of Arizona), 361

\bibitem[]{642} Bastian, N., Gieles, M., Efremov, Yu. N., \& Lamers,
H. J. G. L. M. 2005, A\&A, 443, 79

\bibitem[]{645} Bastian, N., Ercolano, B., Gieles, M., Rosolowsky, E.,
Scheepmaker, R. A., Gutermuth, R., \& Efremov, Yu. 2007, MNRAS, 379,
1302

\bibitem[]{649} Battinelli, P., Efremov, Y., \& Magnier, E.A.
1996, A\&A, 314, 51

\bibitem[]{} Baumgardt, H., Hut, P., \& Heggie, D.C. 2002, MNRAS,
336, 1069

\bibitem[]{652} Boily, C.M., \& Kroupa, P. 2003, MNRAS, 338, 673

\bibitem[]{654} Bonnell, I.A., Bate, M.R., \& Vine, S.G. 2003, MNRAS,
343, 413

\bibitem[]{657} Bonnell, I.A., \& Bate, M.R. 2006, MNRAS, 370, 488

\bibitem[]{659} Brodie, J.P., \& Larsen, S.S. 2002, AJ, 124, 1410

\bibitem[]{661} Burkert, A., Brodie, J., \& Larsen, S. 2005, ApJ, 628,
231

\bibitem[]{664} Clark, P.C. \& Bonnell, I.A. 2005, MNRAS, 361, 2

\bibitem[]{666} Comer\'on, F., Schneider, N., \& Russeil D., 2005,
A\&A, 433, 955

\bibitem[]{669} Dahm, S.E., \& Simon, T. 2005, AJ, 129, 829

\bibitem[]{673} Dickey, J.M., McClure-Griffiths, N.M., Stanimirovic,
S., Gaensler, B.M, \& Green, A.J, 2001, ApJ, 561, 264

\bibitem[]{676} Efremov, Y.N. 1995, AJ, 110, 2757

\bibitem[]{678} Efremov, Y.N., \& Elmegreen, B.G. 1998, MNRAS, 299,
588

\bibitem[]{681} Elias, F., Alfaro, E.J., \& Cabrera-Ca\~no, J. 2006,
AJ, 132, 1052

\bibitem[]{684} Elmegreen, B.G. 2000, ApJ, 530, 277

\bibitem[]{686} Elmegreen, B.G. 2002a, ApJ, 577, 206

\bibitem[]{688} Elmegreen, B.G. 2002b, ApJ, 564, 773

\bibitem[]{690} Elmegreen, B.G. 2005, in The many scales in the
Universe, eds. J. C. del Toro Iniesta, et al. (Dordrecht: Kluwer).
p. 99

\bibitem[]{694} Elmegreen, B.G. 2007, ApJ, 668, in press

\bibitem[]{696} Elmegreen, B. G., Efremov, Y., Pudritz, R. E., \&
Zinnecker, H. 2000, in Protostars and Planets IV, eds. V. Mannings,
A.P. Boss, \& S.S. Russell, (Tucson: Univ. of Arizona Press), p. 179

\bibitem[]{700} Elmegreen, B.G., \& Efremov, Y.N. 1996, ApJ,  466, 802

\bibitem[]{702} Elmegreen, B.G., \& Efremov, Y.N. 1997, ApJ, 480, 235

\bibitem[]{704} Elmegreen, B.G., Elmegreen, D.M., Chandar, R.,
Whitmore, B., Regan, M. 2006, ApJ, 644, 879

\bibitem[]{707} Enoch, M.L. 2006. ApJ, 638, 293

\bibitem[]{709} Feitzinger, J. V., \& Galinski, T. 1987, A\&A, 179,
249

\bibitem[]{} Goodwin, S.P., \& Bastian, N. 2006, MNRAS, 373, 752

\bibitem[]{712} Gomez, M., Hartmann, L., Kenyon, S. J. \& Hewett, R.
1993, AJ, 105, 1927

\bibitem[]{715} Gutermuth, R.A., Megeath, S.T., Pipher, J.L.,
Williams, J.P., Allen, L.E., Myers, P.C., \& Raines, S. N. 2005,
ApJ, 632, 397

\bibitem[]{719} Heitsch, F., Mac Low, M.-M., \& Klessen, R.S. 2001,
ApJ, 547, 280

\bibitem[]{722} Heydari-Malayeri, M., Charmandaris, V., Deharveng, L.,
Rosa, M. R., Schaerer, D., \& Zinnecker, H. 2001, A\&A, 372, 527

\bibitem[]{725} Hunter, D. A. 1999, in IAU Symp. 193, Wolf-Rayet
Phenomena in Massive Stars and Starburst Galaxies, ed. K. A. van der
Hucht, G. Koenigsberger, \& P. R. J. Eenens (San Francisco: ASP),
418

\bibitem[]{730} Ivanov, G. R., Popravko, G., Efremov, Y. N., Tichonov,
N. A., \& Karachentsev, I. D. 1992, A\&AS, 96, 645

\bibitem[]{733} Johnstone, D., Di Francesco, J., \& Kirk, H. 2004,
ApJ, 611, L45

\bibitem[]{736} J{\o}rgensen, J.K. et al. 2006, ApJ, 645, 1246

\bibitem[]{738} J{\o}rgensen, J.K., Johnstone, D., Kirk, H., \& Myers,
P.C. 2007, ApJ, 656, 293

\bibitem[]{741} Kennicutt, R.C., Jr. 1998, ApJ, 498, 541

\bibitem[]{743} Kirk, H., Johnstone, D., \& Di Francesco, J. 2006,
ApJ, 646, 1009

\bibitem[]{746} Klessen, R.S., \& Burkert, A. 2000, ApJS, 128, 287

\bibitem[]{748} Klessen, R.S., Ballesteros-Paredes, J.,
V\'azquez-Semadeni, E., \& Dur\'an-Rojas, C. 2005, ApJ, 620, 786

\bibitem[]{799} Lada, C. J., Margulis, M. \& Dearborn, D., 1984, ApJ, 321, 141

\bibitem[]{751} Lada, C.J., \& Lada, E.A. 2003, ARAA, 41, 57

\bibitem[]{753} Larsen, S.S., \& Richtler, T. 2000, A\&A, 354, 836

\bibitem[]{755} Larsen, S.S., \& Brodie, J.P. 2000, AJ, 120, 2938

\bibitem[]{757} Li, Y., Klessen, R.S., \& Mac Low, M.-M. 2003, ApJ,
592, 975

\bibitem[]{760} Li, P. S., Norman, M,L., Mac Low, M.-M., \& Heitsch,
F. 2004, ApJ, 605, 800

\bibitem[]{763} Li, Z.-Y., Nakamura, F. 2006, ApJ, 640, L187

\bibitem[]{765} Mac Low, M.-M. 2002, in Modes of Star Formation and
the Origin of Field Populations, ASP Conf. Ser. 285, eds. E.K.
Grebel \& W. Brandner, p. 112

\bibitem[]{769} Mac Low, M.-M., \& Klessen, R.S. 2004, RvMP, 76, 125

\bibitem[]{771} McKee, C.F., \& Tan, J.C. 2003, ApJ, 585, 850

\bibitem[]{773} Ma\'iz-Apell\'aniz, J. 2001, ApJ, 563, 151

\bibitem[]{775} Megeath, S.T., et al. 2004 ApJS, 154, 367

\bibitem[]{777} Nanda Kumar, M. S., Kamath, U. S., Davis, C. J. 2004,
MNRAS, 353, 1025

\bibitem[]{780} Nordlund, \AA., \& Padoan, P. 1999, in Interstellar
Turbulence, ed. J. Franco \& A. Carrami\~nana (Cambridge: Cambridge
Univ. Press), 218

\bibitem[]{} Palla, F., \& Stahler, S.W. 2000, ApJ, 540, 255

\bibitem[]{784} Saito, H., Saito, M., Moriguchi, Y., \& Fukui, Y.
2006, PASJ, 58, 343

\bibitem[]{787} Scalo, J.S. 1985, in Protostars and Planets II,  ed.
D.C Black and M. S. Matthews, (Tucson:  Univ. of Arizona Press), p.
201

\bibitem[]{791} Smith, M.D., Gredel, R., Khanzadyan, et al. 2005a,
MmSAI, 76, 247

\bibitem[]{794} Smith, N., Bally, J., Shuping, R. Y., Morris, M., \&
Kassis, M. 2005b, AJ, 130, 1763

\bibitem[]{797} Stanke, T., Smith, M.D., Gredel, R., \& Khanzadyan, T.
2006, A\&A, 447, 609

\bibitem[]{800} St\"utzki, J., Bensch, F., Heithausen, A., Ossenkopf,
V., \& Zielinsky, M. 1998, A\&A, 336, 697

\bibitem[]{853} Tan, J.C. 2007, in Triggered star formation in a turbulent interstellar
medium, IAU Symposium 237, eds. B.G. Elmegreen \& J. Palo\v{s},
(Cambridge: Cambridge Univ Press), p.258

\bibitem[]{803} Testi, L., Sargent, A.I., Olmi, L et al. 2000, ApJ,
540, 53

\bibitem[]{806} Tilley, D.A., \& Pudritz, R.E. 2004, MNRAS, 353, 769

\bibitem[]{808} V\'azquez-Semadeni, E., Ballesteros-Paredes, J. \&
Klessen, R.S. 2003, ApJ, 585, L131

\bibitem[]{811} V\'azquez-Semadeni, E., Kim, J., Shadmehri, M., \&
Ballesteros-Paredes, J. 2005, ApJ, 618, 344

\bibitem[]{814} V\'azquez-Semadeni, E., Kim, J., \&
Ballesteros-Paredes, J. 2005, ApJ, 630, L49

\bibitem[]{817} Verschueren, W. 1990, A\&A, 234, 156

\bibitem[]{819} Wada,~K., \& Norman,~C.~A. 2001, ApJ, 547, 172

\bibitem[]{821} Wada,~K., \& Norman,~C.~A. 2007, ApJ, 660, 276

\bibitem[]{823} Williams, J.P. \& McKee, C.F. 1997, ApJ, 476, 166

\bibitem[]{} Wu, J., Evans, N.J., II., Gao, Y., Solomon, P.M.,
Shirley, Y.L., \& Vanden Bout, P.A. 2005, ApJ, 635, L173

\bibitem[]{825} Zavagno, A., Deharveng, L., Comer\'on, F., Brand, J.,
Massi, F., Caplan, J., Russeil, D. 2006, A\&A, 446, 171

\end{thebibliography}
\end{document}